\documentclass[aps,twocolumn,floatfix,superscriptaddress,showpacs]{revtex4-1}

\usepackage{graphicx}
\usepackage{amsmath}
\usepackage{amssymb}
\usepackage{hyperref}
\usepackage{soul}
\usepackage{dsfont}

\usepackage[]{color}

\newcommand{\e}{\mathrm{e}}

\newcommand{\pdag}{{\phantom{\dagger}}}

\begin{document}
\title{Theory of a 3+1D fractional chiral metal: interacting variant of the Weyl semimetal}

\author{Tobias Meng}
\affiliation{Institut f{\"u}r Theoretische Physik, Technische Universit{\"a}t Dresden, 01062 Dresden, Germany}
\author{Adolfo G. Grushin}
\affiliation{Department of Physics, University of California, Berkeley, CA 94720, USA}
\affiliation{Max-Planck-Institut f{\"u}r Physik komplexer Systeme, 01187 Dresden, Germany}
\author{Kirill Shtengel}
\affiliation{Max-Planck-Institut f{\"u}r Physik komplexer Systeme, 01187 Dresden, Germany}
\affiliation{Department of Physics and Astronomy, University of California at Riverside, Riverside CA 92511, USA}
\author{Jens H. Bardarson}
\affiliation{Max-Planck-Institut f{\"u}r Physik komplexer Systeme, 01187 Dresden, Germany}

\begin{abstract}
Formulating consistent theories describing strongly correlated metallic topological phases is an outstanding problem in condensed matter physics.
In this work we derive a theory defining a fractionalized analogue of the Weyl semimetal state: the fractional chiral metal.
Our approach is to construct a 4+1D quantum Hall insulator by stacking 3+1D Weyl semimetals in a magnetic field.
In a strong enough field the low-energy physics is determined by the lowest Landau level of each Weyl semimetal,
which is highly degenerate and chiral, motivating us to use a coupled-wire approach.
The one-dimensional dispersion of the lowest Landau level allows us to model
the system as a set of degenerate 1+1D quantum wires that can be bosonized in
the presence of electron-electron interactions and coupled such that a gapped
phase is obtained, whose response to an electromagnetic field is given in
terms of a Chern--Simons field theory.
At the boundary of this phase we obtain the field theory of a 3+1D gapless
fractional chiral state, which we show is consistent with a previous theory
for the surface of a 4+1D Chern--Simons theory.
The boundary's response to an external electromagnetic field is determined by a chiral anomaly with a fractionalized coefficient.
We suggest that such anomalous response can be taken as a working definition of a fractionalized strongly correlated analogue of the Weyl semimetal state.
\end{abstract}

\maketitle

%------------------
%-- Introduction --
%------------------

\section{Introduction}
Recent exciting developments in condensed matter physics concern a variety of topological phases.
These are phases that are not classified by broken symmetries and local order parameters~\cite{Qi2011,HasanKane}.
While the term ``topological phases'' was originally used to refer only to topologically ordered gapped phases with long-range entanglement, it is now understood to encompass a broader class of phases.
This includes symmetry-protected topological phases~\cite{Senthil2015}---gapped states with no long-range entanglement that are distinct from the trivial phase only in the presence of certain symmetries---and now frequently even gapless states~\cite{Turner:2013tf,Hosur2013}.

The prototypical gapless topological phase of matter is the nodal semimetal in which nodal points act as sources and sinks of Berry's phase making them topologically stable.
Such states are realized in certain phases of liquid helium and have been discussed extensively in that context~\cite{volovik_book}, while their condensed matter realization in the 3+1D Weyl semimetals is a very recent experimental achievement~\cite{Weng2015,Huang2015,SIN15,Lv2015,XAB15,LvXu2015,YLS15}.
The non-trivial topological structure is induced by spin-orbit coupling.
The surface states corresponding to a pair of Weyl nodes disperse chiraly and result in open Fermi surfaces---the Fermi arcs.
The presence of these exotic surface states is closely related to quantum anomalies which also govern the unusual response of the system to external fields~\cite{Nielsen1983}.
Whether strong electronic correlations can stabilize exotic cousins of nodal phases is the open question that motivates our work.

One fruitful strategy to describe and analyze unconventional 3+1D gapless states of matter is inspired by the fact that
a boundary between gapped phases, one topological and another one not, is expected to be gapless~\footnote{Since topological invariants are quantized, the adiabatic theorem prohibits changing them without closing a gap; one just needs to interpret the coordinate perpendicular to the boundary as a parameter being varied adiabatically. One possible exception to this rule are so-called \emph{quantum doubles} describing a topological state which possesses a pair of invariants of opposite sign; in this case these invariants can ``annihilate'' one another without closing a gap. A toric code provides a simple example of such behavior.}.
Accordingly, the first key idea of this manuscript is to approach the putative fractional Weyl semimetal as a surface state of a higher-dimensional gapped topological state, namely that of a 4+1D fractional quantum Hall state~\cite{ZH01}.
Previous studies have described such a state in terms of Landau levels for Dirac fermions in higher dimensions~\cite{LIY12,LZW13}, quaternions~\cite{YW13}, and
ground state wave functions~\cite{WZ14}.
Our understanding of these phases, however, is still much less advanced in comparison to their 2+1D counterparts.
While some general results, such as the connection between charge fractionalization and a non-trivial ground state degeneracy in a gapped 2+1D system on a torus~\cite{Oshikawa2006}, should carry through in higher dimensions, other aspects of topological order, including for example exotic quasiparticle statistics, are much less transparent.
Describing the 4+1D fractional quantum Hall state, although not the main goal of this study, will therefore be a useful spin-off that adds to the existing body of knowledge of this state.

In this work, we combine the above strategy---approaching a putative fractional 3+1D metallic state as the surface state of a fractional 4+1D quantum Hall state---with a second powerful approach, the so-called coupled-wire construction.
The advantage of the combined strategy is that, unlike previous parton constructions~\cite{WKA14,W15}, it does not postulate fractionalization from the start, allows us to address both
the gapped bulk and the gapless surfaces of a 4+1D quantum Hall state, and remains analytically tractable.
The main idea is to trade the isotropy usually inherent to low-energy topological quantum field theories for the analytical control over electron-electron interactions provided by Luttinger liquid theories describing (coupled) one-dimensional systems.
Starting with the seminal studies \cite{kane_02,TK14}, coupled-wire constructions have been successfully employed to describe a large variety of chiral topological phases in 2+1D \cite{lv12,vb14,kl14,mskl14,so14,ky14,ncmt14,jtl15,SHG15,ms14,sosh15,chm15,skl15,ksl15,oss14,mfsl14,cns15,cs15}, including surfaces of topological 3+1D states \cite{mea15,szt15}, topological superconductors \cite{sbo14,gang_11_14,v14}, and spin liquids \cite{mngt15,gps15,hcgncm16}. First generalizations of the coupled-wire approach to higher dimensions have been discussed in~\cite{v12,m15,so15,INC16}.
In the following, we adapt the coupled-wire construction for the description of topological 4+1D phases, and apply it to a specific class of 4+1D fractional quantum Hall states~\footnote{Different generalizations of the quantum Hall effect to 4+1D exist, which can for instance involve Abelian and non-Abelian gauge fields \cite{ZH01,karabalinair02} -- our constructions is part of the former category}.

In the next section we summarize our main results and discuss the main ideas.
This section is aimed at those readers that are not experts in coupled-wire constructions but are interested in the main ideas behind the calculation.
All the technical details are given in section~\ref{sec:cwc_field} and require some background knowledge; those not interested in these details can safely skip this section and go directly to the discussion in section~\ref{sec:discussion}, where we also discuss the connection to current experimental prospects.
\section{
\label{sec:summary}
Summary of main results}

In this section we discuss the general philosophy and the key ideas of our work; the technical details of our calculation are given in the next section.

The central result of our work is a coupled-wire construction of a 4+1D fractional quantum Hall insulator that has 3+1D fractional chiral metals at its surfaces, and the conjecture of a fractionalized gapless 3+1D phase composed of two fractional chiral metals of opposite chiralities.

The 4+1D quantum Hall insulator we construct has a current density response $j^\mu$ to an external electromagnetic field $A_{\mu}$ given by
\begin{equation}%
\label{eq:5DQHE}
j^{\mu} = \frac{\delta \mathcal{S}_\text{CS}^{(4+1)}}{\delta A_\mu} = C_{2}\dfrac{e^3}{8\pi^2}\epsilon^{\mu\nu\rho\sigma\lambda}\partial_{\nu}A_{\rho}\partial_{\sigma}A_{\lambda},
\end{equation}
where $C_{2} = 1/(2m+1)$ with integer $m\geq0$, $\mu = 0,1,\ldots,4$ and $\epsilon$ is the totally antisymmetric tensor (here and henceforth we use units where $\hbar=c=1$).
The field theory underlying this response is the 4+1D Chern--Simons theory
\begin{align}
\label{eq:CS4+1a}
\mathcal{S}_\text{CS}^{(4+1)}&=\frac{-e^3C_2}{6(2\pi)^2} \int d^5 x \,\epsilon^{\mu\nu\rho\sigma\eta}\,A_\mu\partial_\nu A_\rho \partial_\sigma A_\eta.
\end{align}
According to Eq.~\eqref{eq:5DQHE} a combination of a three dimensional magnetic field $\mathbf{B}$ and an electric field $\mathbf{E}$ both perpendicular to the $x_4$-direction generates a current
\begin{equation}%
\label{eq:j4}
j^{4} = -C_{2}\dfrac{e^3}{4\pi^2}\mathbf{E}\cdot\mathbf{B}
\end{equation}
parallel to the $x_4$-direction \footnote{Here and in the remainder of the paper, we define the three-dimensional vector fields ${\bf E}$ and ${\bf B}$ by their components $E_i$  and $B_i$ ($i=1,2,3$). While not representing the full electromagnetic field strength tensor in (4+1)D, the definition of these fields allows us to explicitly make contact to (3+1)D Weyl semimetals, and underlines the invariance of our construction with respect to rotations around the $x_4$ axis.}.
This result we interpret as the chiral anomaly induced response of a surface 3+1D fractional chiral metal.

This fractional chiral metal interpretation is motivated by an analogy with the edge states in the 2+1D fractional quantum Hall effect.
There, the 2+1D Hall current can be understood as arising from a chiral charge that is pumped from one edge to the other.
The chiral charge is therefore not separately conserved on each edge and the theory describing a given edge is anomalous.
The change in chiral charge is proportional to the electric field inducing the Hall current, with a coefficient that is a fraction
of that obtained in the noninteracting integer case.
Analogously, the response~\eqref{eq:j4} represents the pumping of chiral charge from one anomalous 3+1D metallic surface to another.
For $m>0$, the coefficient of this anomaly is fractionalized with respect to the well-known coefficient of the chiral anomaly of
noninteracting Weyl fermions, which is obtained for $m=0$~\cite{Nielsen1983}.
We thus define the obtained surface state as a fractional chiral metal.

\begin{figure}
  \centering
  \includegraphics[width=\columnwidth]{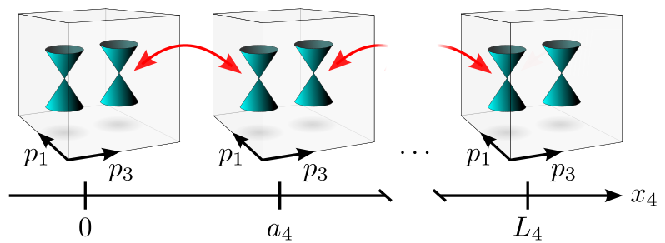}\\\vspace*{0.35cm}\scalebox{2}{$\downarrow$}\hspace*{2.4cm}\scalebox{2}{$\downarrow$}\hspace*{0.45cm}~\\\vspace*{0.5cm}
    \includegraphics[width=\columnwidth]{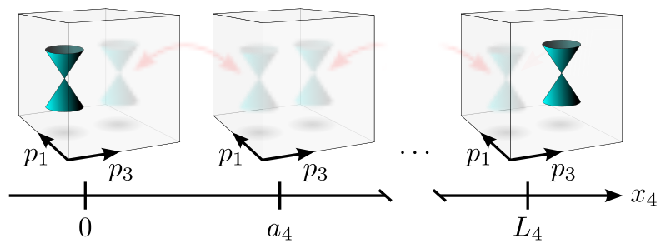}
  \caption{Construction of a 4+1D integer quantum Hall effect from coupled Weyl semimetals. A tunnel coupling between the left-handed node of each Weyl semimetal with the right-handed node of its neighbor at smaller $x_4$, depicted by red arrows, induces a gap for all bulk nodes. In a slab of finite extent $0\leq x_4\leq L_4$, single gapless nodes remain at the 3+1D surfaces at $x_4=0$ and  $x_4=L_4$.
  }
  \label{fig:4dweyl_coupled}
\end{figure}

The fundamental idea of our construction is as follows: A 4+1D quantum Hall state is constructed by regularly stacking 3+1D Weyl semimetals along a fourth spatial direction at $x_4=qa_4$, with $q\in\mathbb{Z}$ and $a_4$ the lattice spacing.
By suitably coupling nodes of opposite chiralities in neighboring Weyl semimetals the bulk is gapped out, as shown schematically in Fig.~\ref{fig:4dweyl_coupled}.
To describe each semimetal, we restrict ourselves to the minimal two-band model of an inversion-symmetric time-reversal-broken Weyl semimetal, which has two Weyl nodes of opposite chirality at the same energy separated in momentum space.
The Hamiltonian describing the Weyl semimetal at $x_4$ is
\begin{align}
H_{0,x_4}= \sum_{{\bf p}}\Psi_{\bf p}^\dagger(x_4)\mathcal{H}_0({\bf p})\Psi_{\bf p}^\pdag(x_4),\label{eq:weyl_nob_1}
\end{align}
where $\mathbf{p}$ is a three dimensional momentum and $\Psi_{\bf p}^\dagger(x_4) = (c^\dagger_{\uparrow,{\bf p}}(x_4),c^\dagger_{\downarrow,{\bf p}}(x_4))$ is a spinor of creation operators for electrons of spin $s \in \{\uparrow,\downarrow$\} and momentum $\mathbf{p}$ at $x_4$.
The spin label more generally denotes the two bands, but for simplicity we always refer to it as spin.
Close to the two Weyl nodes at ${\bf p}^{\chi}=(0,0,\chi b/2)$ with chirality
$\chi = \pm 1$, that is for $|\delta\mathbf{p}	|=|\mathbf{p}-\mathbf{p}^\chi|
\ll b$, the Hamiltonian matrix $\mathcal{H}_0(\mathbf{p})$ is given to lowest
order in $|\delta {\bf p}|/b$ by
\begin{align}
\mathcal{H}_0\big|_{|\delta \mathbf{p}|\ll b} \approx \chi v_F \,\delta {\bf p}\cdot{\boldsymbol\sigma}.
\label{eq:weyl_nob_2}
\end{align}
Here $\boldsymbol\sigma$ is a vector of the three Pauli matrices and $v_F$ is the Fermi velocity.
The detailed form of $\mathcal{H}_0({\bf p})$ away from the Weyl nodes is not important for our construction as long as the separation $b$ is large enough; we comment on the precise conditions where appropriate.

The coupling of the 3+1D Weyl semimetals resulting in a gapped 4+1D quantum Hall state is most transparent in the noninteracting case.
In this case, the right-handed Weyl nodes at $x_4=qa_4$ are coupled to the left-handed nodes at $x_4 = (q+1)a_4$ by tunneling, as depicted by red arrows in Fig.~\ref{fig:4dweyl_coupled}.
Since the two nodes that are so coupled have opposite chirality they can annihilate and gap each other out, resulting in a gapped state.
In a finite slab with $0\leq x_4 \leq L_4$, however, the left-handed node at $x_4=0$ and the right-handed node at $x_4=L_4$ do not have partner nodes to pair up with.
Instead, they form 3+1D chiral gapless surface states that are higher dimensional analogues of the chiral 1+1D edge modes of a 2+1D quantum Hall state.
Like these modes, they escape the fermion doubling theorem~\cite{NielNino81a,NielNino81b,NielNino81c} by the fact that they live on the 3+1D surfaces of a topological 4+1D state.
We therefore identify the gapped state just constructed as an integer 4+1D quantum Hall state.

In the presence of interactions the construction of a gapped state is more involved and we rely on a coupled-wire construction related to that of Kane and collaborators for 2+1D fractional quantum Hall states~\cite{kane_02}.
As in the noninteracting case, we tunnel-couple the left and right handed nodes in neighboring Weyl semimetals, but now the combination of interactions and tunneling leads to several different gapped states, just as in the 2+1D fractional quantum Hall case.
The way this essentially works is that one of the dimension in the Weyl semimetal, say $x_3$, is made into effective one dimensional quantum wires by quenching the kinetic energy along the other $x_1$ and $x_2$ directions.
These effective wires are then coupled through the fourth dimension.
In order to make this calculation controlled, we need the coupling of the wires to be the dominant coupling, such that it leads to a nontrivial gapped phase.
To this end we apply a strong magnetic field $\mathbf{B}=B_3\mathbf{e}_3$, with $\mathbf{e}_3$ the unit vector in the $x_3$ direction, to each of the Weyl semimetals.
This results in the formation of Landau levels that disperse only in $p_3$ and thereby naturally form a basis of quantum wires that can be coupled, see Fig.~\ref{fig:weyl_b_4d}.
At low energies $E \ll 1/l_B$, with the magnetic length defined as
$l_B=1/\sqrt{eB_3}$, each Weyl node can be approximated by its gapless zeroth
Landau level, which is is composed of highly degenerate chiral modes with a
degeneracy factor $N_\text{LL}=L_1 L_2eB_3/2\pi$. Dispersing linearly, these
chiral modes are readily bosonized.
The right-handed chiral modes in one Weyl semimetal are tunnel-coupled to the left-handed chiral modes in the neighboring Weyl semimetal, see Fig.~\ref{fig:weyl_b_4d}.
The inclusion of interaction allows for correlated tunneling in which one particle tunnels between two Weyl semimetals while at the same time $m$ particles in each Weyl semimetal change their chirality from right-handed to left-handed or vice versa, see Fig.~\ref{fig:weyl_B_scattering}.
This is the same multi-particle process as the one of Kane and collaborators leading to the fractional quantum Hall states, and like in that case a current along $x_{4}$ occurs as a result of applying an electric field $\mathbf{E}$ along the wire direction $x_3$.
Here, however, since each of the modes is degenerate the current is also
proportional to the degeneracy factor $N_\text{LL}$ and hence to the magnetic
field $B$.
This is the origin of the chiral anomaly form of the current density in Eq.~\eqref{eq:j4}.

Up to this point we have described how the chiral fractional metal emerges as the boundary state of a 4+1D quantum Hall insulator.
It remains to obtain the field theory description of this boundary state.
To that end we take a manifold finite in the $x_{4}-$direction and impose gauge invariance, to obtain the field theory describing a single surface:
\begin{equation}
\label{eq:edgetheoryintro}
S_{\text{surface}} = \kappa\int_{\partial\Sigma} \partial_{0}\phi d\phi\wedge F
\end{equation}
where $\phi$ is a scalar field defined at that surface, $F=dA$ is the external field strength and $\kappa=\pm e(2m+1)/8\pi^2$ is a constant whose sign defines the chirality of the surface.
This we recognize as the action of a 2+1D quantum Hall effect edge described by $\partial_{0}\phi d \phi$ with an
extra functional dependence $dA$ that, as above, has its origin in the Landau level degeneracy.
Remarkably this result as derived from the coupled-wire construction is consistent with earlier attempts to describe the edge theory of the 4+1D Chern Simons theory~\eqref{eq:CS4+1a} based on the current algebra perspective~\cite{GS94}.

Before going into the details of the calculations, a brief note on coupling scales and renormalization.
The application of a strong magnetic field has the advantage of allowing us to identify the dominant couplings at the microscopic level, leading to a well defined phase.
We further assume that these couplings remain the leading ones under renormalization, and do not attempt a systematic renormalization group analysis of all interaction terms here, as this would take us way beyond the scope of this work.
Instead, in analogy with the canonical discussion of 2+1D fractional quantum Hall states, we simply assume that, if necessary, one can always adjust the values of the microscopic parameters such that the renormalization flow is towards the gapped phases we have identified.

\section{A coupled-wire construction of 4+1D fractional quantum Hall states}\label{sec:cwc_field}

\subsection{A single Weyl node in a magnetic field}\label{subsec:node_in_field}
As discussed in the previous section our coupled-wire construction approach relies on coupling 3+1D Weyl nodes subject to a magnetic field.
Therefore we first recount the physics of a single isotropic 3+1D Weyl node in a magnetic field ${\bf{B}}=B{\bf e}_{3}$, where ${\bf e}_{i}$ denotes the unit vector in $i$-direction (with $i=1,2,3$) \cite{Nielsen1983, ashby_13}.
In the Landau gauge, this magnetic field is associated with the vector potential ${\bf A}=B_3 x_1 {\bf e}_2$.

\begin{figure}
  \centering
\includegraphics[width=\columnwidth]{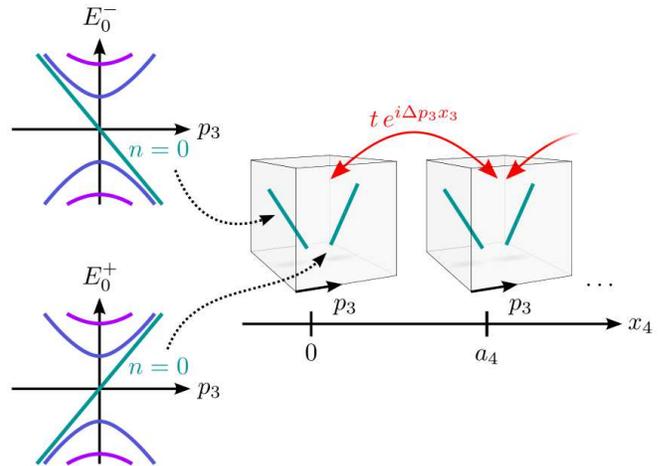}
  \caption{Stacking of Weyl semimetals along the fourth spatial dimension $x_4$ subject to a magnetic field ${\bf B} = B_3 {\bf e}_3$. Each Weyl semimetal contains the Landau levels of a pair of Weyl nodes. The left panels depicts the energy $E^{\chi}_0$ of a Weyl node of chirality $\chi=\pm$ in a magnetic field described by the Hamiltonian $H_{0}^{\chi}$ as a function of the momentum $p_3$ measured with respect to the Weyl node. The Landau levels $n\geq1$ form quadratically dispersing, particle-hole symmetric bands. Each Landau level is macroscopically degenerate with respect to $p_2$. The right panel illustrates the low-energy description of the stacked 4+1D system at finite ${\bf B}$, which reduces to the gapless lowest Landau levels. As discussed in Sec.~\ref{subsec:coupled_wires_4d_b}, a complex hopping $t\,e^{i\Delta p_3 x_3}$ connects neighboring Weyl semimetals.
}
  \label{fig:weyl_b_4d}
\end{figure}

A Weyl node of chirality $\chi$ is described by the first-quantized Hamiltonian
\begin{align}
{H}_{0}^{\chi}=\chi v_F ({\bf p}-e {\bf A})\cdot{\boldsymbol\sigma},\label{eq:weyl_b}
\end{align}
where ${\bf p}=(p_1,p_2,p_3)$ is the three-dimensional momentum measured with respect to the Weyl node, and ${\boldsymbol\sigma}$ is the vector of Pauli matrices.
Choosing the magnetic field such that $e B_3 >0$, with $e$ the electron charge, we define the magnetic length $l_B=1/\sqrt{eB_3}$
and introduce the dimensionless creation and annihilation operators
\begin{subequations}
\begin{align}
a_{p_2}^\pdag &= \frac{1}{\sqrt{2}}\left(\frac{x_1-p_2l_B^2}{l_B} + ip_1l_B\right),\\
a_{p_2}^\dagger &= \frac{1}{\sqrt{2}}\left(\frac{x_1-p_2l_B^2}{l_B} - ip_1l_B\right).
\end{align}\label{eq:bos_ll}
\end{subequations}
These obey the bosonic commutator relation $[{a}_{p_2}^\pdag,{a}_{p_2}^\dagger]=1$.  
In the eigenstates $|p_2\rangle$ of $p_2$ the Hamiltonian takes the form
\begin{align}
\langle p_2|{H}_{0}^{\chi}|p_2'\rangle=\delta_{p_2p_2'}\chi v_F\,\begin{pmatrix} {p}_3&i\sqrt{2} a_{p_2}^\dagger/l_B\\-i\sqrt{2} a_{p_2}^\pdag/l_B&- {p}_3\end{pmatrix}.\label{eq:weyl_ham_a}
\end{align}
Denoting the Landau level quantum number---the integer eigenvalues of $a^\dagger_{p_2}a_{p_2}$---by $n$ the spectrum of~\eqref{eq:weyl_ham_a} comprises particle-hole symmetric bands with dispersion $E_{0,n>0}^{\chi}(p_3)= \pm\chi \sqrt{v_F^2p_3^2+2 n/l_B^2}$ and a chiral linearly dispersing lowest Landau level $E_{0,n=0}^{\chi}(p_3) =  \chi v_F p_3$, as illustrated in Fig.~\ref{fig:weyl_b_4d}.
Since the bands are independent of the momentum eigenvalue $p_2$ it labels the degenerate states inside each Landau level whose number is
\begin{align}
N_\text{LL}=\frac{L_1 L_2B_3}{2\pi/e},
\end{align}
where $L_i$ is the length of the system in $i$-direction.

When the magnetic field ${B_3}$ is sufficiently large for all energy scales of interest to be smaller than $v_F/l_B$, the low-energy physics is determined by the gapless lowest Landau level only.
Thus a single Weyl node can be approximated by a macroscopically degenerate set of right- or left-moving chiral electrons with a one-dimensional dispersion $E_{0,n=0}^{\chi}(p_3)$.
The eigenvectors of this lowest Landau level take the form $(|n=0,p_2\rangle, 0)$ and are therefore spin-polarized.

\subsection{The coupled-wire Hamiltonian and correlated tunneling processes}
\label{subsec:coupled_wires_4d_b}
We now construct the full 4+1D system by stacking individual Weyl semimetals in the limit of strong magnetic field ${\bf B}=B_3 {\bf e}_3$
where we can restrict our model to the gapless lowest Landau levels.
We choose the stacking to be along the additional discrete spatial dimension $x_4=q a_4$ with  $q\in\mathbb{Z}$, see Fig.~\ref{fig:weyl_b_4d}.
The associated Hamiltonian reads
\begin{align}
H= H_{0}+H_\text{tun} + H_\text{int},\label{eq:b_4d_basic_ham}
\end{align}
where $H_{0}$ describes the individual Weyl semimetals, and $H_\text{tun}$ and
$H_\text{int}$ encode, respectively, tunneling terms between them and the
electron--electron interactions.

Let us start by detailing $H_0$.
We take the Fermi energy in each of the semimetals to reside at the Weyl nodes,
which  are located at momentum ${\bf p}^{\chi}=(0,0,\chi b/2)$.
Sec.~\ref{subsec:node_in_field} then implies that the low-energy form of $H_0$
is captured by linearly dispersing right- and left-moving modes dispersing only
with $p_3$. Adapting the standard low-energy treatment of one-dimensional
systems \cite{giamarchi_book}, we approximate $H_0$ by a model with unbounded
linear dispersions,
\begin{align}
H_{0}=\sum_{x_4,p_2}& \int dx_3\Bigl(R_{p_2}^\dagger(x_3,x_4)\,(-i v_F \partial_3) \, R_{p_2}^\pdag(x_3,x_4)\nonumber\\&+ L_{p_2}^\dagger(x_3,x_4)\,i v_F \partial_3 \, L_{p_2}^\pdag(x_3,x_4)\Bigr).
\label{eq:Hzero}
\end{align}
Here, the chiral modes $R_{p_2}^\pdag$  and $L_{p_2}^\pdag$ precisely
correspond to the linearly dispersing low-energy excitations in the lowest
Landau levels shown in Fig.~\ref{fig:weyl_b_4d}.
As customary in bosonization, the translation from the initial operators
$c_{p_2}^\pdag(x_3,x_4)$ and $c_{p_2}^\dagger(x_3,x_4)$, creating and
annihilating an electron of momenta $p_2$ in the lowest Landau level at
position $(x_3,x_{4})$, to the chiral low-energy modes is via
\begin{align}
c_{p_2}^\pdag(x_3,x_4) = e^{-i x_3b/2}L_{p_2}^\pdag(x_3,x_4)+ e^{i x_3b/2} R_{p_2}^\pdag(x_3,x_4).\label{eq:operators}
\end{align}

Next, we address the explicit form of the tunneling term $H_\text{tun}$.
At vanishing magnetic field, we require the tunneling to preserve the momenta $p_1$ and $p_2$.
In the presence of a strong magnetic field, this translates into a conservation of the Landau level index $n$ and the momentum $p_2$.
The electron momentum $p_3$, however, is allowed to be shifted by the tunneling, by an amount $\Delta p_3$ that depends on the precise state to be generated.
Physically, a finite momentum shift demands that the electrons couple to a vector potential with an $x_4$-component of $\mathcal{A}_4= \Delta p_3 x_3/e a_4$.
This can be achieved with a complex hopping whose phase $\Delta p_3 x_3 = e\int_{qa_4}^{(q+1)a_{4}}dx_4 \mathcal{A}_{4}$ relates to the vector potential via the Peierls substitution.
Such a Peierls phase of the complex hopping is indeed equivalent to a momentum shift $\Delta p_3$ for an electron tunneling from $x_4=q a_4$ to $x_4=(q+1)a_4$.
Denoting the tunneling strength by $t$, we thus find that the projection of the tunneling Hamiltonian to the lowest Landau levels reads
\begin{align}
H_\text{tun}&=\sum_{q,p_2,p_3}t\,c^{\dagger}_{p_2,p_3+\Delta p_3}((q+1)a_4)\,c_{p_2,p_3}(qa_4)+\text{H.c.}
\end{align}
where $c_{p_2,p_3}(x_4)$ denotes the Fourier transform of $c_{p_2}^\pdag(x_3,x_4)$ with respect to the third coordinate.

The Hamiltonian $H_\text{int}$, finally, describes electron-electron
interactions, whose presence is a crucial ingredient to fractional quantum Hall
states.
The screening of long range interactions by the large density of states of the gapless lowest Landau level motivates us to neglect non-local interactions.
Since in addition the wave function of an electron with degeneracy index $p_2$ is proportional to a Gaussian centered at $x_{1}=p_2l_B^2$, the largest contribution to the local interaction involves electrons with the same $p_2$.
We thus specialize to couplings $\rho_{p_2}(x_3,x_4) \,\rho_{p_2}(x_3+a_3,x_4)$ between the densities $\rho_{p_2} = {c}_{p_2}^\dagger {c}_{p_2}^\pdag$ of electrons with identical $p_2$ at the closest possible coordinates $(x_3,x_4)$ and $(x_3+ a_3,x_4)$, where $a_3$ is the lattice constant along $x_3$.
Using $c_{p_2}^\pdag(x_3+ a_3,x_4) \approx c_{p_2}^\pdag(x_3 ,x_4)+a_3 \partial_3 c_{p_2}^\pdag(x_3 ,x_4)$, we obtain the interaction Hamiltonian as
\begin{multline}
H_\text{int} =\int d x_3 \sum_{p_2,x_4} U {c}_{p_2}^\dagger(x_3,x_4)(\partial_3{c}_{p_2}^\dagger(x_3,x_4)) \\
\times(\partial_3 {c}_{p_2}^\pdag(x_3,x_4)){c}_{p_2}^\pdag(x_3,x_4).\label{eq:special_int_ham}
\end{multline}
The effects of further interaction processes not relevant to our discussion are briefly addressed in Sec.~\ref{subsec:interactions} below.

\begin{figure}
 \centering
  \includegraphics[width=\columnwidth]{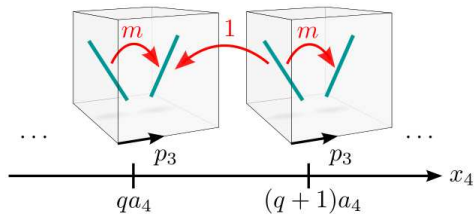}
  \caption{Correlated tunnelings between neighboring Weyl semimetals leading to fractional quantum Hall states. While an electron hops from the left-handed Weyl-node at $x_4=(q+1)a_4$ to the right-handed node at $qa_4$, $m$ electrons are scattered from the left-handed node to the right-handed node in both Weyl semimetals connected by the tunneling. This correlated process is indicated by the arrows, whose labels indicate the number of electrons transported along the respective arrow.
  }
  \label{fig:weyl_B_scattering}
\end{figure}

In 2+1D coupled-wire constructions, topologically ordered states are generated by correlated tunnelings of electrons between wires; these are processes in which an electron tunnels from one wire to a neighboring wire, while simultaneously a number of additional electrons in both wires are backscattered \cite{kane_02}.
We generalize this class of processes to 4+1D by analyzing the correlated tunnelings  depicted in Fig.~\ref{fig:weyl_B_scattering},
in which an electron tunnels from the Weyl semimetal at $x_4=(q+1) a_4$ to the neighboring semimetal at $x_4=qa_4$, while at the same time $m$ electrons are backscattered between the Weyl nodes of both semimetals.

\begin{figure}
 \centering
  \includegraphics[width=\columnwidth]{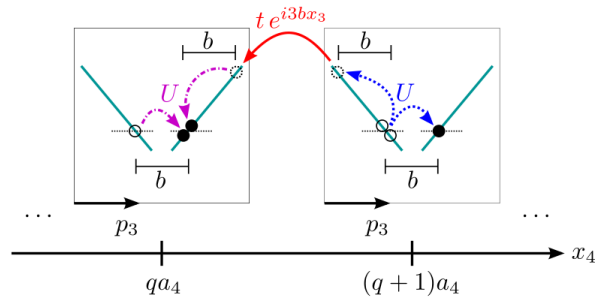}
  \caption{Generation of the correlated scattering $\sim tU^2$ corresponding to $m=1$ in Fig.~\ref{fig:weyl_B_scattering}.  Empty circles depict initial states of the electrons, filled circles correspond to their final states, and the dotted circles indicate the intermediate virtual states of the hopping electron. The dotted horizontal lines depict the Fermi level (and thus the energy of the Weyl nodes).}
  \label{fig:weyl_B_scattering_bulk}
\end{figure}

Microscopically, the correlated tunnelings in Fig.~\ref{fig:weyl_B_scattering}
are obtained from the Hamiltonian in Eq.~\eqref{eq:b_4d_basic_ham} by treating
$H_\text{tun}$ and  $H_\text{int}$ as perturbations to the decoupled Weyl
semimetals described by $H_0$.
Let us start by illustrating the derivation in the simplest case $m=1$.
This process is generated by the combined perturbative expansion of
$H_\text{tun}$ and $H_\text{int}$ to first order in the tunneling $t$ and
second order in $U$, see Fig.~\ref{fig:weyl_B_scattering_bulk}.
As shown by the dotted arrow, the local interaction $U$ first causes two electrons at the left-handed node in the Weyl semimetal at $x_4=(q+1) a_4$ to scatter off each other with a $p_3$ momentum transfer of $b$;
one electron thus ends up at the right-handed node, while the other electron occupies an intermediate high-energy state at momentum $p_3 = -3b/2$.
The high-energy electron then hops, as indicated by the solid arrow, to the Weyl semimetal at $x_4=qa_4$ by virtue of the complex tunneling $t\,e^{-i\Delta p_3 x_3}$.
It thereby acquires a momentum shift $-\Delta p_3$ and ends up at $p_3 = -3b/2-\Delta p_3$.
Finally, the interaction $U$ mediates a scattering, depicted by the dash-dotted arrow, between the tunneling electron and an electron that is initially close to the left-handed node at $x_4 = q a_4$.
Because the total correlated tunneling process has to conserve energy (and momentum), this second electron should be scattered to the right-handed Weyl node at $x_4 = q a_4$ (or any other Weyl node for that matter, but these other processes do not generate the fractional quantum Hall states we are interested in; see also Sec.~\ref{subsec:interactions}).
The second scattering process is thus also associated with a momentum transfer of $b$.
The tunneling electron has thereafter acquired a total $p_3$-momentum shift of
$\Delta p_3^\text{tot}=-2b-\Delta p_3$.
The process shown in Fig.~\ref{fig:weyl_B_scattering_bulk}, which stabilizes a fractional quantum Hall state, is a correlated tunneling that transfers the hopping electron from the left-handed Weyl node at $x_4=(q+1) a_4$ to the right-handed node at $x_4=q a_4$.
This requires the total momentum shift for the tunneling electron to be $\Delta p_3^{\rm{tot}}=+b$, and thus fixes $\Delta p_3 = -3b$ for this particular process to conserve momentum.

To obtain a low-energy description of this process, we  integrate out the high-energy intermediate states of the hopping electron.
We then obtain an effective three-particle interaction that annihilates two left movers with identical quantum number $p_2$ at the left-handed node at $x_4=(q+1) a_4$, and creates two right movers at the right-handed node at $x_4=q a_4$.
Due to their fermionic character, these right and left movers cannot be at the same position, but need to be slightly displaced.
Since we generate the process using the interaction in Eq.~\eqref{eq:special_int_ham} involving derivatives $\partial_3$, this is indeed the case.
For $\Delta p_3=-3b$, we find the  low-energy Hamiltonian of the correlated tunneling shown in Fig.~\ref{fig:weyl_B_scattering_bulk} to read

\begin{align}
H_{\text{tun}}^{(2)}&\sim tU^{2} \sum_{p_2,q}\int dx_3 \, R_{p_2}^\dagger(x_3,qa_4) L_{p_2}^\pdag\left(x_3,(q+1)a_4\right)\nonumber\\
&\times R_{p_2}^\dagger\left(x_3,(q+1)a_4\right)\left[\partial_{3}L_{p_2}^\pdag\left(x_3,(q+1)a_4\right)\right]\nonumber\\
&\times \left[\partial_3 R_{p_2}^\dagger(x_3,qa_4)\right]L_{p_2}^\pdag(x_3,qa_4)+\rm{H.c.}.\label{eq:corrrel_hooping_ferm_op}
\end{align}
For $\Delta p_3\neq -3b$, when the process depicted in
Fig.~\ref{fig:weyl_B_scattering_bulk} does not conserve momentum,
Eq.~\eqref{eq:corrrel_hooping_ferm_op} acquires additional oscillating factors
$\text{exp}(\pm i x_3(\Delta p_3+3b))$ that suppress the scattering.
This is analogous to the observation that a 2+1D Laughlin state only exists at specific filling fractions (i.e., specific strengths of the applied magnetic field), and that the 2+1D coupled-wire construction of these states involves a momentum shift proportional to the applied field~\cite{kane_02}.

The analogy to the 2+1D case, where Laughlin states exist for many filling factors, immediately suggests that there should be a number of momentum shifts $\Delta p_3\neq -3b$ at which other correlated tunnelings conserve momentum in our 4+1D system.
This is indeed the case for $\Delta p_3 = -(2m+1) b$ with $m\in\mathds{Z}^+$, when processes depicted in Fig.~\ref{fig:weyl_B_scattering} with a backscattering of $m$ electrons between the Weyl nodes become resonant.
These higher-order correlated tunnelings are generated in first order in
$H_\text{tun}$ and $2m$-th order in $H_\text{int}$, and are thus $\sim t
U^{2m}$.
The first $m$ interaction processes now scatter $m$ electrons in the Weyl semimetals $x_4 = (q+1)a_4$ from the left-handed the the right-handed node, thereby transferring a momentum of $- m b$ to one other electron that is initially in the vicinity of the left-handed node.
This latter electron is thus pushed to a momentum $p_3=-(m+1/2)b$.
It then tunnels to $x_4=q a_4$, and thereby acquires a momentum shift of $\Delta p_3=-(2m+1)b$, which puts it at a momentum $p_3=(m+1/2)b$.
In the final $m$ interaction processes,  $m$ electrons are scattered from the left-handed node to the right-handed node at $x_4=qa_4$ while transferring a momentum of $- m b$ to the electron that has tunneled.
This latter electron consequently end up at momentum $p_3=b/2$, i.e., at the right-handed node.
As a subtlety, we remark that all intermediate states of the tunneling electron should be at high energies (within the lowest Landau level approximation), and thus far away from the Weyl nodes.
As a result, the possible values of $m$ for which our construction is valid are constrained by the periodicity of the Brillouin zone to satisfy  $b/2+mb < \pi/a_3$ and by the lowest Landau level approximation to $mb \ll 1/l_B$.

The low-energy Hamiltonian describing these higher-order correlated tunnelings can be obtained in analogy to Eq.~\eqref{eq:corrrel_hooping_ferm_op} by integrating out the intermediate high-energy states of the tunneling electron.
We then obtain a scattering process involving $2m+1$ low-energy electrons that
includes the annihilation of $m+1$ left moving electrons with identical $p_3$
at $x_4=(q+1) a_4$, as well as the creation of $m+1$ right moving electrons at
$x_4=qa_4$.
Correspondingly, the low-energy Hamiltonian again contains derivatives with
respect to $x_3$ that account for the small spatial displacements of the
individual electrons.
We obtain
\begin{multline}
H_{\text{tun}}^{(2m)}\sim tU^{2m}\, \sum_{p_2,q}\int dx_3 \, R_{p_2}^\dagger\left(x_3,qa_4\right)^\pdag
L_{p_2}^\pdag\left(x_3,(q+1)a_4\right)\\
\times \prod_{j=1}^m\left[\partial_3^{j-1}R_{p_2}^\dagger\left(x_3,(q+1)a_4\right)\right]
\left[\partial_3^j L_{p_2}^\pdag(x_3,(q+1)a_4)\right]\\
\times \left[\partial_3^jR_{p_2}^\dagger(x_3,qa_4)\right]\left[\partial_3^{j-1}L_{p_2}^\pdag(x_3,qa_4)\right]+\rm{H.c.}.\label{eq:corrrel_hooping_ferm}
\end{multline}
Just as Eq.~\eqref{eq:corrrel_hooping_ferm_op}, this Hamiltonian is only valid
if the resonance condition $\Delta p_3=-(2m+1) b$ is met.
All correlated tunnelings with $m'\neq m$ violate momentum conservation, and are suppressed by oscillating factors.

\subsection{Other interaction processes}\label{subsec:interactions}
\begin{figure}
 \centering
  \includegraphics[width=\columnwidth]{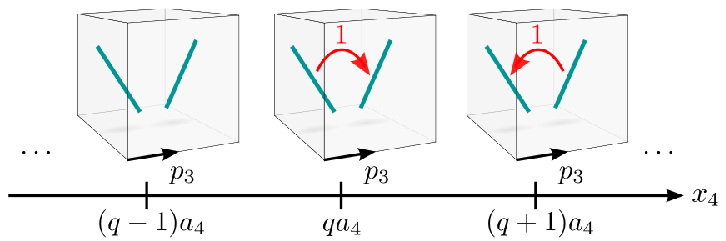}\\[0.5cm]
    \includegraphics[width=\columnwidth]{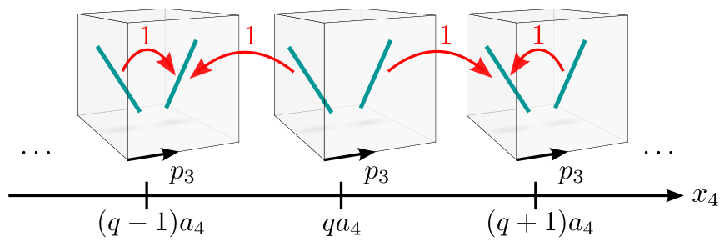}
  \caption{Two of the possible competing interaction processes that would lead to a charge-density-wave order (top panel), and a 1/3-Laughlin-crystal-type of order (bottom panel) \cite{kane_02}.}
  \label{fig:weyl_B_scattering_competing}
\end{figure}

The interaction in Eq.~\eqref{eq:special_int_ham} is not of the most generic form, but is rather optimized to explore a specific set of states in our 4+1D system.
Namely, we are interested in the states generated by the couplings shown in Fig.~\ref{fig:weyl_B_scattering}.
Those are the 4+1D analogues of the correlated scatterings generating Laughlin states in a 2+1D coupled-wire system \cite{kane_02}.
We disregard possible competing states generated by Eq.~\eqref{eq:special_int_ham}, or by a more general interaction.
This includes locked charge density waves in neighboring Weyl semimetals
generated by backscattering interactions, or a 1/3-Laughlin-crystal type of
order \cite{kane_02}, see Fig.~\ref{fig:weyl_B_scattering_competing}.

In general, any of the couplings present in a given system may determine the low-energy physics.
Technically, this happens if the respective term has a large coupling constant, while all other terms have small coupling constants.
The hierarchy of coupling constants is either due to a fine-tuning of their bare values, or can be generated by their RG flow.
The structure of the flow is ultimately determined by the interaction itself, which can again be tuned to favor a particular coupling.

Since the remainder of this paper aims at characterizing the phases generated
by the correlated tunnelings depicted in Fig.~\ref{fig:weyl_B_scattering},
%.
%
we neglect all other interaction processes by assuming that the system
parameters have been be adjusted accordingly.
Note, however, that we analyze a topological state of matter that is not symmetry-protected.
The topological response given in Eq.~\eqref{eq:j4} is thus insensitive to the
addition of other interactions provided these remain sufficiently small and do
not alter the RG scaling of the correlated tunneling process depicted in
Fig.~\ref{fig:weyl_B_scattering} by preventing it from being the most relevant
term.
This philosophy has already proven very useful for the exploration of topological phases in 2+1D coupled-wire constructions.
A full RG analysis of all sine-Gordon terms that could possibly be present in a coupled-wire system is, however, to date lacking even for these much simpler systems.
It constitutes an important open problem for the field in general, and is beyond the scope of the present work.

\subsection{Bosonization}\label{subsec:bosonization}
We bosonize the chiral modes using the standard prescription \cite{giamarchi_book}

\begin{equation}
  r_{p_2}^{}(x_3,x_4)=
  \frac{U_{rp_2x_4}^{}}{\sqrt{2\pi\alpha}}\,\e^{-i\Phi_{rp_2}(x_3,x_4)},
  \label{eq:quasi_part_bos}
\end{equation}
where $U_{rp_2x_4}^{}$ is a Klein factor, $\alpha^{-1}$ denotes a high momentum
cutoff, and $r_{p_2}=R_{p_{2}},L_{r_{p_2}}$ is a compact notation
for the operators introduced in Eq.~\eqref{eq:Hzero}.
The chiral bosonized fields obey

\begin{align}
  \bigl[\Phi_{r{p}_2}(x_3,x_4),\Phi_{r'{p}_2'}(x_3',x_4')\bigr] &=\delta_{rr'}\delta_{{p}_2{p}_2'}\delta_{x_4x_4'}\nonumber\\& \times i\pi \hat{r}\,\text{sgn}(x_3-x_3'),\label{eq:bos_comm_b}
\end{align}
where the scalar $\hat{r}$ takes the value $\hat{r}=+1$ ($\hat{r}=-1$) for $r=R$ ($r=L$).
In the Luttinger liquid Hamiltonian describing the low-energy physics of the
bosonized modes $\Phi_{rp_2}$, we retain the bosonized version of $H_0$, the
correlated tunnelings $H_\text{tun}^{(2m)}$, and interactions $\sim \rho_{rp_2}
\rho_{r'p_2}$ between the densities of the chiral modes.
In bosonization, the latter are given by $\rho_{rp_2}=r_{p_2}^\dagger r_{p_2}^\pdag = -\hat{r}\partial_{x_3} \Phi_{r p_2}/2\pi$.
We neglect all further interactions, including in particular the ones shown in Fig.~\ref{fig:weyl_B_scattering_competing} and other interaction processes that turn right movers into left movers.
Upon bosonization, these scatterings give rise to sine-Gordon terms that
compete with the correlated tunnelings $H_\text{tun}^{(2m)}$; as for the terms
depicted in Fig.~\ref{fig:weyl_B_scattering_competing}, we consider system
parameters such that all of these sine-Gordon terms are irrelevant for our
system.

The gapless motion described by $H_0$ and the chiral density-density
interactions can be combined into a quadratic bosonized Hamiltonian
$H_0^\text{bos}$ reading

\begin{align}
\label{eq:H0Hint}
&H_0^\text{bos}= \sum_{r,q,q',p_2}\int dx_3\nonumber\\& \times(\partial_{x_3}{\Phi}_{r p_2}(x_3,qa_4))^T\,V_{p_2,q,q'}\,(\partial_{x_3}{\Phi}_{r p_2}(x_3,q'a_4)),
\end{align}
where

\begin{align}
\label{eq:V}
V_{p_2,q,q'}&=\frac{v_F}{4\pi}\,\delta_{qq'}\mathds{1}+\tilde{U}_{p_2,q,q'}
\end{align}
depends on the Fermi velocity $v_F$ and the density-density interactions $\tilde{U}_{p_2,q,q'}$ between the different bosonized modes \cite{giamarchi_book}.
In order to bosonize the tunnelings  in Eq.~\eqref{eq:corrrel_hooping_ferm}, it is useful to introduce new fields
\begin{subequations}
\begin{align}
\widetilde{\Phi}_{L{p}_2}(x_3,x_4) &= (m+1)\Phi_{Lp_2}(x_3,x_4)-m\Phi_{R{p}_2}(x_3,x_4),\\
\widetilde{\Phi}_{R{p}_2}(x_3,x_4) &= (m+1)\Phi_{R{p}_2}(x_3,x_4)-m\Phi_{L{p}_2}(x_3,x_4),
\end{align}\label{eq:basis_trafo_b}
\end{subequations}
which obey
\begin{multline}
[\widetilde{\Phi}_{r{p}_2}(x_3,x_4),\widetilde{\Phi}_{r'{p}_2'}(x_3')]=\delta_{rr'}\delta_{{p}_2{p}_2'}\delta_{x_4x_4'}\\
\times(2m+1)\,i\pi \hat{r}\,\text{sgn}(x_3-x_3').\label{eq:supp_comm_b}
\end{multline}
This definition, together with Eq.~\eqref{eq:quasi_part_bos}, allows us to cast the leading terms in the operator product expansion of Eq.~\eqref{eq:corrrel_hooping_ferm} into the form
\begin{multline}
H_{\text{tun}}^{(2m)}\sim tU^{2m}\sum_{q,{p}_2} \int dx_3 \\
\times\cos \Bigl(\widetilde{\Phi}_{L{p}_2}(x_3,(q+1)a_4)-\widetilde{\Phi}_{R{p}_2}(x_3,qa_4)\Bigr).\label{eq:tun_weyl_4d_bos_up_b}
\end{multline}
Because the argument of each sine-Gordon term in Eq.~\eqref{eq:tun_weyl_4d_bos_up_b} commutes with itself at different positions, each term can order individually by pinning its argument to one of its minima.
Since the arguments also commute between different sine-Gordon terms, all of them can order simultaneously.
This fully gaps the bulk.
If the system is finite along $x_4$ and has 3+1D surfaces at $x_4=0$ and $x_4=L_4$, two sets of surface modes  remain gapless. These are the modes $\widetilde{\Phi}_{L{p}_2}(x_3,0)$ and $\widetilde{\Phi}_{R{p}_2}(x_3,L_4)$ which simply do not have a partner mode to pair up with and thus do not appear in any of the sine-Gordon terms.

\subsection{Field theory}\label{sec:field_theory}
In order to show that the bulk of the gapped state obtained above behaves as a fractional 4+1D quantum Hall state,
we calculate its response to an external electromagnetic field in a quantum field theory representation using the action formalism.
This further allows us to relate the boundary modes to a 3+1D chiral anomaly with a fractional coefficient as compared with the noninteracting case.

In previous sections we established that a stack of 3+1D Weyl semimetals is
described by a Hamiltonian of the form $H=H_{0}+H_\text{tun}^{(2m)}$, where
$H_{0}$ is the bosonized free theory of Eq.~\eqref{eq:H0Hint} and
$H_{\text{tun}}^{(2m)}$  encodes the relevant correlated tunneling terms of
Eq.~\eqref{eq:tun_weyl_4d_bos_up_b}.
In order to explicitly derive the response of this system to an external electromagnetic field we closely follow Ref.~\onlinecite{SHG15}, in which a Chern-Simon theory of 2+1D fractional topological insulators was obtained, while highlighting the differences.

The starting point of the derivation is to implement, within the coupled-wire construction, minimal coupling of the electromagnetic field $A_{\mu}$ to the fermionic current $j_{\mu}$ through $j^{\mu}A_{\mu}$ where the summation over $\mu=0,1,2,3,4$ is implied.
It is technically convenient to treat the different components separately.
The $\mu =1,2$ components were already included in the construction of the Landau levels, and the theory respects gauge invariance in these coordinates through the Landau level degeneracy prefactor $N_{LL}$, which is proportional to the (gauge invariant) magnetic field $B_{3}=\partial_{1}A_{2}-\partial_{2}A_{1}$.
The rest of the components are obtained by demanding that our theory is invariant under the gauge transformation:
\begin{subequations}
\label{eq:gauge_trafo_ferm}
\begin{align}
A_\alpha &\to A_\alpha+ \partial_\alpha \xi,  &\alpha=0,3, \\
A_4 &\to A_4+ \partial_4 \xi, & \\
c_{p_2} &\to e^{ie \xi}\,c_{p_2}, 	
\end{align}
\end{subequations}
with an analogous relation for $c^{\dagger}_{p_{2}}$.
Here and for the remainder of this section indices $\alpha,\beta$ imply summation over $0,3$ only while $\mu,\nu$ imply summation over all indices.
We suppress in the notation the explicit dependence of all fields on $(x_0,x_3,x_4)$; the dependence on $x_1$ and $x_2$ is encoded in the Landau level quantum numbers $p_{2}$ and $n$.
Within the realm of bosonized fields the gauge transformation~\eqref{eq:gauge_trafo_ferm} translates, via Eq.~\eqref{eq:quasi_part_bos}, into
\begin{align}
\Phi_{r,{p}_2} &\to \Phi_{r,{p}_2}-e \xi,\\
\widetilde{\Phi}_{r,{p}_2} &\to \widetilde{\Phi}_{r,{p}_2}-e \xi .
\end{align}
The gauge invariant generalization of the tunneling term Eq.~\eqref{eq:tun_weyl_4d_bos_up_b}, which contains the electromagnetic response of the 4+1D state along the $x_{4}$ direction, then takes the action form
\begin{align}\label{eq:tun_weyl_4d_bos_up_b_gauge}
&S_{\text{1}}[\Phi,A_4]\sim t U^{2m}\sum_{q,p_2} \int dx_{0}dx_3\\& \cos \Bigl(\widetilde{\Phi}_{L,p_2}((q+1)a_4)-\widetilde{\Phi}_{R,p_2}(qa_4)+ea_4 A_{4}(qa_4) \Bigr).\nonumber
\end{align}
We have assumed $A_{4}$ is a smooth function of $x_4$ on scales $l\gg a_4$, where $a_4$ is the lattice constant in the discrete fourth dimension.
The remaining components involve $j_{\alpha}=(\rho,j_{3})$ with $\rho$ the particle density and $j_3$ the current density along the $x_3$ direction.
To write the corresponding coupling $j^\alpha A_\alpha$ in terms of the $\widetilde{\Phi}$ fields, we note that the one-dimensional particle density of electrons at $x_4$ is given by
\begin{align}
\rho_{} &= \sum_{{p}_2}\frac{1}{2\pi}\partial_3 ({\Phi}_{L,{p}_2}-{\Phi}_{R,{p}_2})\nonumber\\
&= \sum_{{p}_2}\frac{1}{2\pi(2m+1)}\partial_3 (\widetilde{\Phi}_{L,{p}_2}-\widetilde{\Phi}_{R,{p}_2}).\label{eq:charge}
\end{align}
This implies that a $2\pi$ kink in one of the sine-Gordon terms of Eq.~\eqref{eq:tun_weyl_4d_bos_up_b} carries a charge of $e/(2m+1)$. The (one-dimensional) particle current density $j_3$, on the other hand, is related to the particle density by the continuity equation
\begin{align}
\label{eq:conserv}
\partial_{\alpha}j^{\alpha}=0.
\end{align}
Using this continuity equation and the density expression~\eqref{eq:charge} we obtain the electromagnetic response
\begin{align}
\label{eq:SEMwires}
\mathcal{S}_\text{2}&[\Phi,A_{\alpha}]=e\sum_{x_4} \int dx_0 \,dx_3\,j^{\alpha}_{} A_{\alpha}\\
&=\sum_{{p}_2,x_4} \frac{e}{2\pi(2m+1)}\int dx_0\, dx_3\,\epsilon^{\alpha\beta}\partial_\alpha (\widetilde{\Phi}_{R{p}_2}-\widetilde{\Phi}_{L{p}_2})\,A_{\beta},\nonumber
\end{align}
with $\epsilon^{\alpha\beta}$ the totally antisymmetric tensor and summation over $\alpha,\beta \in \{0,3\}$ is implied.
By combining the above action components, Eqs.~\eqref{eq:tun_weyl_4d_bos_up_b_gauge} and \eqref{eq:SEMwires}, with the $A_\mu$ independent kinetic term $\mathcal{S}_0[\Phi]$ obtained from Eq.~\eqref{eq:H0Hint}, the full electromagnetic response of the system finally takes the form
\begin{align}
\label{eq:b_action}
\mathcal{S}=\mathcal{S}_0[\Phi]+\mathcal{S}_\text{1}[\Phi,A_4]+\mathcal{S}_\text{2}[\Phi, A_{\alpha}].
\end{align}
In the next step towards showing that the response~\eqref{eq:b_action} reduces to that of a 4+1D fractional quantum Hall effect, we use the fact that the cosines in Eq.~\eqref{eq:tun_weyl_4d_bos_up_b_gauge} pin their arguments to one of their minima when these terms become relevant.
This in turn fixes the difference between left and right chiral fields to
\begin{align}
\label{eq:gauge_pinning}
\widetilde{\Phi}_{R,p_2}(qa_4)-\widetilde{\Phi}_{L,p_2}((q+1)a_4)=ea_4 A_{4}(qa_4).
\end{align}
With this strong-coupling condition we write the action $\mathcal{S}_2$ in Eq.~\eqref{eq:SEMwires} in terms of $A_{4}$ and $A_{\alpha}$ only
\begin{align}
\label{eq:almost_gauge_inv_form}
\mathcal{S}_\text{2}[A_{4},A_{\alpha}]=\sum_{p_2}\frac{-e^2}{2\pi(2m+1)}\int dx_0dx_{3}dx_4\,\epsilon^{\alpha\beta}  A_{4}\,\partial_\alpha A_{\beta}.
\end{align}
Importantly, the summands are now independent of $p_2$ which allows us to
replace the sum $\sum_{p_{2}}$ with the Landau level degeneracy $N_\text{LL}=
eB_3 L_{1}L_{2}/2\pi$.
This action yields the current density
\begin{equation}%
\label{eq:j4e3b3}
j^{4} = \frac{\delta \mathcal{S}_{\rm}}{\delta A_4} = -\frac{e^3}{4\pi^2(2m+1)}B_3E_{3}.
\end{equation}
One immediately identifies this as the response~\eqref{eq:j4} of a 4+1D quantum Hall effect to an external electromagnetic field that satisfies $\mathbf{B}=B_3\mathbf{e}_3$ and $\mathbf{E}=E_{3}\mathbf{e}_3$.

To finally connect this response to a 4+1D Chern--Simons term, we express the
Landau level degeneracy as
\begin{equation}
\label{eq:NLL}
N_\text{LL}= \frac{B_3 L_1 L_2}{2\pi/e}  = \int dx_1 dx_2 \frac{1}{2\pi/e} \epsilon^{\gamma\delta}\partial_{\gamma}A_\delta,
\end{equation}
where  $\gamma,\delta\in \{1,2\}$.
With this, the action $S_\text{2}$ in Eq.~\eqref{eq:almost_gauge_inv_form}
becomes
\begin{align}
\mathcal{S}_\text{2}[A_{\mu}]&=\frac{-e^3}{(2\pi)^2(2m+1)}\int d^5x\,\epsilon^{\alpha\beta}\epsilon^{\gamma\delta}  A_{4}\,(\partial_\alpha A_{\beta})\,(\partial_{\gamma} A_\delta),
\label{eq:4d_s_em_inc}
\end{align}
where $\alpha,\beta \in \{0,3\}$ and $\gamma,\delta\in \{1,2\}$.
This action can be interpreted as a part of the isotropic fractionalized
Chern--Simons action
\begin{align}
\label{eq:CS4+1}
\mathcal{S}_\text{CS}^{(4+1)}[A_{\mu}]&=\frac{-e^3}{6(2\pi)^2(2m+1)} \int d^5 x \,\epsilon^{\mu\nu\rho\sigma\eta}\,A_\mu\partial_\nu A_\rho \partial_\sigma A_\eta.
\end{align}
This action exactly generates the response given by~\eqref{eq:5DQHE} with a fractional $C_{2}=1/(2m+1)$, with the current density along $x_{4}$ taking the form
\begin{equation}%
\label{eq:j4_maintext}
j^{4} = -\frac{e^3}{4\pi^2(2m+1)}\mathbf{B}\cdot\mathbf{E},
\end{equation}
reproduced in Eq.~\eqref{eq:j4}.

Due to its intrinsic anisotropy, the coupled-wire construction only recovers the part of Eq.~\eqref{eq:j4_maintext} given by Eq.~\eqref{eq:j4e3b3}.
Note, however, that we are free to redefine the direction of the external magnetic field as well as
the direction along and perpendicular to the wires.
Had we chosen other directions, we would have separately obtained all the parts
that compose the Chern--Simons action~\eqref{eq:CS4+1}.
Another argument for the correctness of the isotropic Chern--Simons functional
form of \eqref{eq:CS4+1} is that the complete isotropy is required from gauge
invariance in the bulk.
A way to understand this is by analogy with emergent 2+1D Chern--Simons terms
in \emph{non-relativistic} field theories (dimensionality is irrelevant for the
present argument).
If Lorentz invariance is preserved, one expects the functional form $\epsilon^{\mu\nu\rho} a_{\mu} \partial_{\nu} a_{\rho}$ for a generic bosonic field $a_{\mu}$, with all terms having the same coefficient.
The absence of Lorentz invariance naively would lead one to expect different prefactors for $a_{0}\partial_{i}a_{j}$ and $a_{i}\partial_{0}a_{j}$ but this statement is incorrect;
a Chern--Simons field theory is only gauge invariant if all of its terms are
present with the same coefficient \cite{Zee2010}.

Considering all the above, we conclude that Eq.~\eqref{eq:j4_maintext} represents the isotropic current response of the gapped state found in our coupled-wire construction.
This is the response of a fractional 4+1 D quantum Hall insulator to an external electromagnetic field, as advertised by Eq.~\eqref{eq:j4}.

\subsection{Surface theory: Fractional chiral metal}
Having obtained a bulk description of the 4+1D fractional quantum Hall state, we now characterize its boundary modes---the chiral fractional metal.

Our object of interest is the 3+1D theory at the surface of the 4+1D quantum Hall insulator.
From our findings in the previous section, see
Eqs.~\eqref{eq:almost_gauge_inv_form} and~\eqref{eq:NLL}, the 3+1D surface
state should be constructed from $N_\text{LL}$ copies of the edge of a 2+1D
Chern--Simons theory, each copy labelled by $p_2$.
A useful and general way to access such surface theories is to describe the quantum Hall states with an effective theory of conserved current operators $J^{\mu}$ in terms of bosonic fields $b_{\rho\sigma\cdots}$ (the number of indices is determined by dimensionality) so that
$J^{\mu}=\epsilon^{\mu\nu\rho\sigma\cdots}\partial_{\nu}b_{\rho\sigma\cdots}$
satisfies $\partial_{\mu}J^{\mu}=0$ by construction~\cite{Wen2007,Zee2010}.
In our case, the effective action is a sum of 2+1D actions for each copy labelled by $p_{2}$
\begin{eqnarray}
\label{eq:2+1bftheory}
\mathcal{S}_\text{eff} [b^{(p_2)}_{\mu},A_{\mu}] &=& \int d^3x\sum_{p_2} \mathcal{L}_{\text{eff}}[b^{(p_2)}_{\mu},A_{\mu}];\\
\nonumber
\mathcal{L}_{\text{eff}}[b^{(p_2)}_{\mu},A_{\mu}]&=& \dfrac{2m+1}{4\pi} \epsilon^{\mu\nu\rho}b^{(p_2)}_{\mu}\partial_{\nu}b^{(p_2)}_{\rho}\\
&-&\dfrac{e}{2\pi} \epsilon^{\mu\nu\rho}A_{\mu}\partial_{\nu}b^{(p_2)}_{\rho}. %+\cdots
\label{eq:2+1bftheory2}
\end{eqnarray}
This theory recovers the characteristic 2+1D Chern--Simons term after
integrating out the $N_{LL}$ gauge fields $b^{(p_2)}_{\mu}$
\begin{align}
\label{eq:2+1CS}
\mathcal{S}_\text{CS}[A_{\mu}]=\frac{e^2}{4\pi(2m+1)}\sum_{p_{2}}\int d^3x\,\epsilon^{\mu\nu\rho}  A_{\mu}\,\partial_\nu A_{\rho},
\end{align}
resulting in the fractional Hall conductivity $\sigma_{H}=\frac{e^2}{(2m+1)h}$ for each $p_{2}$.
The Chern--Simons theory Eq.~\eqref{eq:almost_gauge_inv_form} in its isotropic
form is obtained by summing over $p_{2}$ which results in an overall  prefactor
of $N_\text{LL}$ in front of the 2+1D Chern--Simons theory~\eqref{eq:2+1CS}
provided $\mu,\nu,\rho \in \{0,1,4\}$.
For notational simplicity and until otherwise stated, we now drop the label $p_2$.
To write down the edge theory of each 2+1D dimensional `slice' defined in the
$(x_{0},x_{1},x_{4})$ coordinate space, consider an infinite strip
$\Sigma=\mathbb{R}^2\times[0,L_4]$ of length $L_4$ in the $x_{4}$ direction.
Firstly, it is possible to check~\cite{SHG15} that gauge invariance at the boundary $\partial\Sigma$ requires that $b_{\mu}|_{\partial\Sigma}=\partial_{\mu}\phi$
where $\phi$ is a scalar field, which we will interpret physically shortly.
By using this constraint and partial integration of the action~\eqref{eq:2+1bftheory} we can rewrite its first term as (see Appendix \ref{app:Ssurf} for details)
\begin{eqnarray}
\nonumber
S_\text{eff}[b_{\mu},0] &=&\dfrac{(2m+1)}{2\pi}\int_{\Sigma} [b_{4}(\partial_{0}b_3-\partial_{3}b_{0})+ b_3\partial_{4}b_{0}]\\
\nonumber
&+&\dfrac{(2m+1)}{4\pi}\int_{\partial\Sigma} [\partial_{0}\Phi_{R}\partial_{3}\Phi_{R}- \partial_{0}\Phi_{L}\partial_{3}\Phi_{L}],\\
\label{eq:Ssurf}
\end{eqnarray}
where we have identified the two chiral bosonized fields defined in \eqref{eq:quasi_part_bos} with the scalar field $\phi$ at the edges such that
$\Phi_{R}:=\phi(x_0,x_3,x_{4}=L_{4})$ and $\Phi_{L}:=\phi(x_0,x_3,x_{4}=0)$.
In other words we have associated a physical meaning to $\phi$: it is the bosonic scalar field that represents the
two chiral modes at the boundary.

The surface theory is completed by the non-universal Hamiltonian defined by the first term in Eq.~\eqref{eq:V}.
Adding it to Eq.~\eqref{eq:Ssurf}, the full surface theory is
\begin{eqnarray}
\nonumber
S_{\text{surface}}[\Phi_{R,L}]&=& \dfrac{2m+1}{4\pi}N_\text{LL} \int_{\partial\Sigma} [\partial_{0}\Phi_{L}\partial_{3}\Phi_{L}-v_{F}(\partial_{3}\Phi_{L})^2\\
\label{eq:surface}
&-&(\partial_{0}\Phi_{R}\partial_{3}\Phi_{R}+v_{F}(\partial_{3}\Phi_{R})^2)].
\end{eqnarray}
Note that since each $p_{2}$ copy, described by the pair of fields $\Phi^{(p_2)}_{L,R}$,  is decoupled from the rest, they all contribute equally to the path integral.
Therefore, the summation over $p_{2}$ results in the $N_{LL}$ prefactor reinstated above, and the field theory can be written in terms of a single copy described by the pair $\Phi_{R,L}$.
This action describes two chiral modes propagating in opposite directions.
To understand this, recall that from \eqref{eq:charge} the right and left moving densities are
$\rho_{R,L} \propto \partial_{x}\Phi_{R,L}$.
The equations of motion of $\Phi_{R,L}$ are thus the continuity equations for the two chiral surface fluids, namely
\begin{eqnarray}
\partial_{0}\rho_{R,L}+(-1)^{\hat{r}}v_{F}\partial_{z}\rho_{R,L}=0
\end{eqnarray}
where $\hat{r}=\pm1$ for $R,L$ respectively.

To summarize our findings so far, we have found that the coupled-wire
construction leads to a description of a fractional 4+1D quantum Hall effect
composed of $N_\text{LL}$ copies of 2+1D Chern--Simons theories.
All copies together lead to a surface theory that is physically $N_\text{LL}$
copies of fractional chiral edge modes of a 2+1D fractional quantum Hall
effect.
We define these surface modes as the (critical) 3+1D fractional chiral metal theory we were after.

To further support our conclusion, we now relate the field theory emerging at
the boundary of a 4+1D Chern--Simons theory found by earlier work~\cite{GS94}
to our findings based on the coupled-wire construction presented above.
Based on a current algebra analysis of the 4+1D Chern--Simons field theory in
closed boundaries, Gupta and Stern~\cite{GS94} wrote a consistent field theory
of the boundary modes, with action
\begin{equation}
\label{eq:edgetheory}
S_{\phi F} = \kappa\int_{\partial\Sigma} \partial_{0}\phi d\phi\wedge F
\end{equation}
where $\phi$ is a scalar field and $F=dA$ is a 2-form that is closed and non-dynamical in $\partial\Sigma$.
The action~\eqref{eq:edgetheory} defines the coupling of a scalar field to an external divergenceless field $\tilde{B}^{i}\equiv \epsilon^{ijk}F_{jk}$.

To make the connection with the surface theory explicit note that, rewriting the $N_{LL}$ prefactor using~\eqref{eq:NLL} we can write Eq.~\eqref{eq:Ssurf} in the functional form
\begin{equation}
\label{eq:edgetheory2}
S_\text{surface}[\Phi,A_{\alpha}] = -\hat{r}\dfrac{(2m+1)e}{8\pi^2}\int_{\partial\Sigma} \partial_{0}\Phi_{r} \partial_{3}\Phi_{r} \epsilon_{3\beta\alpha}\partial_{\beta}A_{\alpha}
\end{equation}
for a single surface that we write using the compact notation $\hat{r}=+1,-1$ for $\Phi_r=\Phi_{L},\Phi_{R}$ respectively.
Comparing ~\eqref{eq:edgetheory} with ~\eqref{eq:edgetheory2} we find that the former is the generalization of the latter if
we identify the divergenceless field $\tilde{B}_{i}$ in Gupta and Stern's construction with the external magnetic field ($\tilde{B}_{i}=B_{i}$) that creates the 3+1D Weyl semimetal Landau levels.
Such an identification provides a physically meaningful interpretation of the more mathematical construction of Ref.~\cite{GS94}.
In addition, this analysis fixes the coefficient to $\kappa=e(2m+1)/8\pi^2$.
The structure of the surface theory~\eqref{eq:edgetheory2} shows that the result emerging from
the coupled-wire construction is consistent with an independent analysis
based on the current algebra analysis of the 4+1D Chern Simons field theory.
We therefore identify~\eqref{eq:edgetheory2} as the action describing the theory of a fractional chiral metal in 3+1D.

\begin{figure}
  \centering
  \includegraphics[width=\columnwidth]{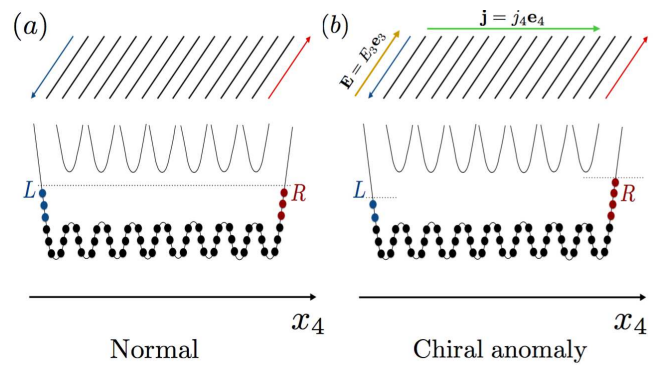}
  \caption{(a) Upper panel: Wire construction of the a 2+1D quantum Hall effect.
  Lower panel: Each wire has a quadratic dispersion relation that is gapped by inter-wire tunneling terms at
  integer filling, depicted by the dashed line.
  The lowest band is composed of boundary left and right movers denoted by blue and red filled circles respectively, distinct from bulk states (black circles).
  (b) When an electric field is applied, left movers are pumped to right movers through the bulk states.
  The non-conservation of left and right chiralities, manifested through the distinct left and right chemical potential depicted as
  dashed lines, is the $1+1$D chiral anomaly.
  Our construction can be thought of as $N_\text{LL}$ copies of this state, each contributing with its own chiral anomaly.
  The surface states of the 4+1D quantum Hall effect constructed in this way defines the chiral metal at the boundary (see main text).
  }
  \label{fig:anomaly}
\end{figure}

We conclude by justifying why the fractional metal is actually ``chiral".
We start building our argument for each 2+1D quantum Hall effect copy by
recalling that the Hall response is intimately related to the chiral anomaly of the edge states in 1+1D.
In Fig.~\ref{fig:anomaly}(a) we show the pictorial representation of a quantum Hall (or Chern insulator) state as obtained in the coupled-wire construction~\cite{kane_02}.
Upon applying an electric field in the $x _{3}$ direction $\mathbf{E}=E_{3}\mathbf{e}_{3}$ [Fig.~\ref{fig:anomaly} (b)] the Fermi momentum $k_{F}$ of the left (right) chiral mode is shifted downwards (upwards).
This creates a deficit of left movers in favor of an excess of right movers and the difference between their densities is proportional to the electric field.
This phenomenon, the chiral anomaly~\cite{Nielsen1983} in 1+1D pumps electrons
through the bottom of the lowest band, generating a Hall current along $x_{4}$.
In our 4+1D construction the magnetic field $\mathbf{B}=B_3\mathbf{e}_{3}$
splits the system into 1+1D wires that are $N_\text{LL}$ degenerate.
Each wire experiences a (fractional) chiral anomaly in 1+1D.
Since there are $N_{LL}$ wires, the total pumped chiral charge adds up to the 3+1D chiral anomaly given by Eq.~\eqref{eq:j4}.

\section{Discussion and conclusions}
\label{sec:discussion}
In this section we motivate and conjecture the existence of 3+1D systems
exhibiting the physics we discussed thus far in terms of 4+1D quantum Hall edge
states, contextualize our results within existing experimental proposals to
simulate higher dimensions, discuss the principal advantages of our
construction, and describe alternative approaches and open problems.

\subsection{General remarks on potential realizations of 3+1D fractional chiral metals}
An evident and fundamental question is whether the surface (or surfaces) of the fractional quantum Hall
states in 4+1D described in this work can lead to new insights into strongly-correlated phases in 3+1D.
While individual surface states of 4+1D integer quantum Hall states cannot
exist in a purely 3+1D system, two of its surface states (i.e., two Weyl nodes)
of opposite chiralities can be combined into a 3+1D Weyl semimetal.
Therefore, it is natural to  expect that the same holds true for the 4+1D
fractional quantum Hall states we have constructed here, which leads us to
conjecture the existence of a gapless fractionalized 3+1D phase whose
properties can be understood by combining two fractional chiral metals of
opposite chiralities in a 3+1D system.

A simplistic toy model that realizes such a fractionalized generalization of a Weyl semimetal starts from a minimalistic version of our construction in Sec.~\ref{sec:cwc_field}: two Weyl semimetals stacked along a fourth direction.
As elaborated further below, the two sites along the fourth direction can be achieved through an internal two-level degree of freedom of a 3+1D system.
Upon adding the same couplings between the Landau levels of the two stacked Weyl semimetals as we do for the 4+1D bulk system in Sec.~\ref{sec:cwc_field}, one obtains the 3+1D analogue of a fractional helical Luttinger liquid~\cite{oss14,mfsl14}.
This indicates that the current flowing in such a collection of fractional helical Luttinger liquids is indeed only a fraction of the one expected for a Weyl semimetal.

Although a more realistic 3+1D model remains to be found, our construction
provides a clear fingerprint of the proposed fractionalized gapless 3+1D phase:
its response to an applied electromagnetic field is that of
a Weyl semimetal with a fractionalized prefactor.
Such a signature should be accessible by standard experimental
probes (see e.g.~\cite{Son2013,Parameswaran2014a}), and, even more importantly,
its topological origin guarantees its independence of system-specific details.\\

Furthermore, it is interesting to observe that recent progress brings our
seemingly purely academic considerations for the bulk 4+1D fractional quantum
Hall states closer to real experimental setups.
In particular, research on 4+1D integer quantum Hall states has lately also been fueled by experimentally accessible proposals for quantum simulators of five dimensional space-time systems.
For instance, the connectivity of a collection of coupled superconducting qubits and resonators determines the effective
dimensionality of a given array and thus can simulate extra- or even fractal-dimensional Ising models~\cite{TAN10}.
Alternatively, extra dimensions can be encoded via additional degrees of freedom in optical lattices~\cite{BCL12}, a strategy recently
exploited to propose an experimental realization of the 4+1D quantum Hall effect using ultra-cold atoms~\cite{PZO15}.
In an orthogonal approach, it was argued that quasicrystalline materials may
encompass higher dimensional topological structure~\cite{KLR12} (see also
Ref.~\cite{MBB13} for a discussion of topological equivalence to regular
crystals).
In particular, the energy gaps of a 2+1D quasicrystal are characterized by the second Chern number~\cite{YRZ13}, a topological invariant
associated to the 4+1D quantum Hall effect in class A~\cite{SRF08}.
In a nutshell, the long-range quasiperiodic order in two dimensions can be mapped to two additional degrees of freedom that, when added to the two-dimensional momenta, construct the higher dimensional invariant.
Finally, bi-volume photonic lattices---two photonic lattices coupled by evanescent modes---can encode the extra degrees of freedom needed to
simulate a synthetic extra dimension, and have been proposed to access higher dimensional solitons~\cite{JB13}.
In addition, the different modes of photonic resonator arrays can be used to mimic synthetic gauge fields and extra dimensions for photons~\cite{SKP15},
that in turn can also engineer higher dimensional quantum Hall physics~\cite{OPG15}.

Among these proposals, allocating the extra dimension in additional degrees of freedom in optical lattices is perhaps the most appealing route to access interacting 4+1D quantum Hall physics~\cite{BCL12,PZO15}.
Multi-orbital cold atomic set-ups will naturally include residual interactions that could be helpful in driving a strongly correlated phase.
Although the particular form of the interaction needed to drive our state is yet to be determined, cold atomic set-ups may allow to realize the coupled-wire construction discussed in Sec.~\ref{sec:cwc_field}.
This motivates the further investigation of more practical issues such as the effect of the trapping potential~\cite{GDD13,DG13,ALS14,SGM15}, the role of heating~\cite{LDM14,LDM14b,DR14,PPH15},  as well as how to physically access the boundary state.
\\

Finally, it is worth mentioning that 
the construction presented here relies on the presence of a strong magnetic field $B_3$, which masks the isotropic nature of Weyl nodes.
Hence, even though the chiral anomaly and the subsequent fractional response
are manifest in our construction, the fact that we cannot reach the isotropic
point makes us reluctant to call the resulting phase a fractional Weyl
semimetal. This motivates us to propose the name chiral fractional metal.
Still, we believe that fractional chiral metals with $m>0$ are
smoothly connected to their zero magnetic field limit, in which case they
should be described by an isotropic theory. This is certainly true for the
non-interacting $m=0$ case corresponding to the usual Weyl nodes.

The chiral character of the surface states is physically most
apparent for $m=0$, for which the surface theory
\eqref{eq:surface} more precisely describes the Landau levels of chiral Weyl nodes. By analogy, the fractional case $m>0$ corresponds to the Landau levels of the fractional chiral metal.
One might hope to gain further insights into the fractional case by
refermionizing the system at fine-tuned values of the interactions, by analogy
to Luther--Emery points \cite{giamarchi_book,lutheremery}, but we have not been able to find such a point in our
construction.

\subsection{Conclusions}
The main advancement of our work is twofold: first, it adds to the small but steadily increasing body of knowledge on the nature of the 4+1D quantum Hall effect; second, it is an attempt
towards a consistent definition of a 3+1D fractional chiral metallic phase.

Regarding the first point, our construction of the 4+1D quantum Hall effect has two advantages: i) it naturally incorporates by construction the chiral anomaly at the boundary and ii) it provides a natural way to write down the boundary field theory, one of our main results.
Although we leave it here as an open question, it is plausible to expect that our construction can be connected to earlier studies of these states~\cite{ZH01,LIY12,ETB12,YW13,LZW13,WZ14}.
In particular, the parton construction is a natural framework for the $1/(2m+1)$ Laughlin like states we obtain.

Related to the second point, the fractional chiral metal is fundamentally a new state of matter.
The description and classification of strongly correlated phases is generally a hard open question to address.
Due to the unbounded growth of the correlation length in metals, the search is oftentimes restricted to gapped phases.
This context highlights how the construction presented here is useful: going to higher dimension, we have been able
to define what a chiral fractional metal is, through its response to an external electromagnetic field.
We have done so by defining this phase as the state that responds to an external electromagnetic field not preserving chirality.
The technical key achievement is to calculate this response from the gapped bulk action of a higher-dimensional topological state, and to rely on the bulk-boundary correspondence to identify our result with the response of the strongly interacting surface states.
This way, we find that the fractional chiral metal existing at each edge of the 4+1D fractional quantum Hall effect
pumps chiral charge to the opposite boundary through the bulk, only now the magnitude of this phenomenon is
a fixed fraction of that corresponding to the non-interacting case of Weyl semimetals.

An important ingredient of our coupled-wire construction is the selection of the wire axis by the applied magnetic field, which quenches the kinetic energy in the $(x_1,x_2)$-plane.
Alternatively, one could stack Weyl semimetals with strongly anisotropic Weyl nodes, in which the velocity in a given direction, $x_3$ for instance, is much larger than the rest, e.g., the $(x_1,x_2)$-plane.
Treating the motion in this plane as a perturbation to the dominantly one-dimensional motion along $x_3$, we have checked that correlated tunnelings analogue to the ones depicted in Fig.~\ref{fig:weyl_B_scattering} can lead to a bulk gapped state with gapless edge modes.
In the integer quantum Hall case, the edge modes recover a single anisotropic Weyl node per surface once the weak dispersion in the $(x_1,x_2)$-plane is taken into account.
In the fractional case, the edge modes do not correspond to free fermions, but still exhibit an anisotropic three-dimensional dispersion.

Finally, our work opens a number of future directions.
The coupled-wire construction can be naturally generalized to incorporate
non-Abelian states~\cite{TK14} and it would be appealing to define topological
defects with anyonic statistics in chiral 3+1D metals~\cite{INC16}.
What is missing at this point is a microscopic Hamiltonian that can support a
fractional chiral metal as its ground state.
Our construction has defined such a state, thus making the first step toward
realization of this novel phase of matter;
we expect our work to trigger further research in strongly correlated metallic topological phases.\\

\section{Acknowledgements}

AGG thanks B. A. Bernevig, T. Neupert and C. Chamon for useful insights in the
early stages of this work and support from the European Commission under the
Marie Curie Programme. TM acknowledges discussions with M. Vojta, S. Rachel and
C. Repellin, and support by the Helmholtz association through VI-521, and the
DFG through SFB 1143. JHB acknowledges ERC Starting Grant No.\ 679722. This
research was supported in part by the National Science Foundation under Grants
No.\ NSF DMR-1411359 (KS) and  No.\ NSF PHY11-25915 (JHB).

\appendix

\section{\label{app:Ssurf}Explicit derivation of \eqref{eq:Ssurf}}

Here we show how to explicitly obtain \eqref{eq:Ssurf} starting from the action
\begin{eqnarray}
\label{eq:2+1bftheoryapp}
\mathcal{S}_{\text{eff}}= C_{1}  \int d^{3}x\epsilon^{\mu\nu\rho}b_{\mu}\partial_{\nu}b_{\rho}
\end{eqnarray}
with $C_{1}= \frac{2m+1}{4\pi}$ that corresponds to the first term in the Lagrangian specified in Eq.~\eqref{eq:2+1bftheory2}.
 This term can be rewritten as
 \begin{eqnarray}
 \nonumber
 S_\text{eff }&=& C_1 \int d^{3}x b_{0}(\partial_{3}b_{4}-\partial_{4}b_3)\\
\label{eq:CSinter}
&+&b_3(\partial_{4}b_{0}-\partial_{0}b_{4})+b_{4}(\partial_{0}b_3-\partial_{3}b_{0})\\
\nonumber
\label{eq:CSf}
&=& 2C_1 \int d^{3}x [b_{4}(\partial_{0}b_3-\partial_{3}b_{0})]\\
&+& C_1  \int d^{3}x [b_3\partial_{4}b_{0}-b_{0}\partial_{4}b_{x}]
\end{eqnarray}
In the last step we have integrated by parts the first and fourth term in \eqref{eq:CSinter} and neglected surface contributions.
We are allowed to do so since we are interested in the manifold $\Sigma=\mathbb{R}^2\times\left\lbrace[0,L_4]\right\rbrace$,
i.e. an infinite strip of length $L_4$ in the $x_{4}$ direction.
Note that the anisotropic coupled-wire construction misses the last two terms in the bulk theory in order to recover the full Chern Simons theory~\cite{SHG15}.
Integrating by parts the last term of Eq.~\eqref{eq:CSf} and noting that in this case we must keep the boundary terms
we obtain
\begin{eqnarray}
\nonumber
S_\text{eff}&=&2C_{1}\int_{\Sigma} d^{3}x [b_{4}(\partial_{0}b_3-\partial_{3}b_{0})+ b_3\partial_{4}b_{0}]\\
\nonumber
&+& C_{1}\int_{\Sigma} dx_{0} dx_{3} (b_{0}b_3)\big|^{x_{4}=L_{4}}_{x_{4}=0} \\
\nonumber
&=&\int_{\Sigma} d^{3}x 2C_{1} [b_{4}(\partial_{0}b_3-\partial_{3}b_{0})+ b_3\partial_{4}b_{0}]\\
\label{eq:Ssurfapp}
&+&C_{1}\int_{\partial\Sigma} dx_{0} dx_{3} [\partial_{0}\Phi_{R}\partial_{3}\Phi_{R}- \partial_{0}\Phi_{L}\partial_{3}\Phi_{L}],
\end{eqnarray}
where in the last step we have i) used the gauge invariant constraint $b_{\mu}|_{\partial\Sigma}=\partial_{\mu}\phi$ and ii)
defined the two chiral bosonized fields at the boundary as $\Phi_{R}:=\phi(x_{0},x_{3},x_{4}=L_{4})$ and $\Phi_{L}:=\phi(x_{0},x_{3},x_{4}=0)$
as indicated in the main text (c.f. Eq.~\eqref{eq:quasi_part_bos}).

\end{document}